\documentclass[multphys,ind]{svmult}
\usepackage{amssymb}
\usepackage{latexsym}
\usepackage{epic, eepic}
\usepackage{graphicx} \DeclareGraphicsExtensions{.ps}

\newcommand{\tC}{\widetilde C}
\newcommand{\ep}{\epsilon}
\newcommand{\CC}{{\mathbb C}}

\newcommand{\NN}{{\mathbb N}}
\newcommand{\RR}{{\mathbb R}}

\def\t2{{\mathbb T}^2}
\newcommand{\ZZ}{{\mathbb Z}}

\newcommand{\F}{{\mathcal F}}
\newcommand{\Hh}{{\mathcal H}}
\newcommand{\Oo}{{\mathcal O}}

\newcommand{\ra}{\rangle}

\newcommand{\hn}{\Hh_{N}}
\newcommand{\defeq}{\stackrel{\rm{def}}{=}}
\def\hto0{\stackrel{h\to 0}{\longrightarrow}}
\def\Nto8{\stackrel{N\to \infty}{\longrightarrow}}

\newcommand{\bequ}{\begin{equation}}
\newcommand{\set}[1]{\left\{\,#1\,\right\}}

\title*{Fractal Weyl Law for Open Chaotic Maps}
\author{St\'ephane Nonnenmacher}
\institute{Service de Physique Th\'eorique, 
CEA/DSM/PhT, Unit\'e de recherche associ\'ee au CNRS,
CEA/Saclay,\\
91191 Gif-sur-Yvette, France\\
{\tt snonnenmacher@cea.fr}}

\begin{document}

\maketitle


\section{Introduction}

We summarize our work
in collaboration with Maciej Zworski \cite{nz}, on the semiclassical density
of resonances for a quantum open system, in the case when the associated classical
dynamics is uniformly hyperbolic, and the set of {\em trapped} trajectories is 
{\em fractal repeller}. 
The system we
consider is not a Hamiltonian flow, but rather a ``symplectic map with a hole'' 
on a compact phase space (the 2-torus). Such a map can be considered
as a model for the Poincar\'e section associated with a scattering 
Hamiltonian on $\RR^2$, at some positive
energy; the ``hole''
represents the points which never return to the Poincar\'e section, that is, which are scattered to 
infinity. We then quantize this open map, obtaining a sequence of subunitary operators, the eigenvalues
of which are interpreted as resonances.

We are especially interested in the asymptotic density of ``long-living resonances'',
representing metastable states which decay in a
time bounded away from zero (as opposed to ``short resonances'', associated with states decaying 
instantaneously). Our results (both numerical and analytical) support the conjectured
{\em fractal Weyl law}, according to which the number of long-living resonances scales as 
$\hbar^{-d}$, where $d$ is the (partial) fractal dimension of the trapped set.

\subsection{Generalities on resonances}
A Hamiltonian dynamical system (say, $H(q,p)=p^2+V(q)$ on $\RR^{2n}$)
is said to be ``closed'' at the energy $E$ when the energy surface $\Sigma_E$ is 
a compact subset of the phase space. The associated quantum operator $H_\hbar=-\hbar^2 \Delta +V(q)$
then admits discrete spectrum near the energy $E$ (for small enough $\hbar$).
Furthermore, if
$E$ is nondegenerate (meaning that the flow of $H$ has no fixed point on $\Sigma_E$), then
the semiclassical density of eigenvalues is given by the celebrated Weyl's law \cite{Ivr}:
\begin{equation}
\label{eq:weyl-closed} 
\# \set{\mathrm{Spec}( H ) \cap [ E- \delta, E + \delta ]}  =  
\frac{1}{ ( 2 \pi \hbar )^n } \int \! \! \int_{ |H(q,p) - E | < \delta } 
\D q\, \D p + \Oo ( \hbar^{1-n} )\,.
\end{equation}
This formula connects the density of quantum eigenvalues with the geometry of the classical
energy surface $\Sigma_E$. It shows that the number of resonances in an interval of type
$[E+C\hbar,E-C\hbar]$ is of order $\Oo(\hbar^{1-n})$. Intuitively, this Weyl law means that
one quantum state is associated with each phase space cell of volume $(2\pi\hbar)^n$.

\medskip

When $\Sigma_E$ is non-compact, or even of infinite volume, 
the spectral properties of $H_\hbar$ are different. Consider the
case of a scattering
situation, when the potential $V(q)$ is of compact support: for any $E>0$, $\Sigma_E$ 
is unbounded, and $H_\hbar$
admits absolutely continuous spectrum on $[0,\infty)$. However, 
one can meromorphically continue the resolvent $(z-H_\hbar)^{-1}$ across the real axis from the
upper half-plane into the lower half-plane. 
In general, this continuation will have discrete poles $\{z_j=E_j-\I\gamma_j\}$ 
with ``widths'' $\gamma_j > 0$, which are the {\em resonances} of $H_\hbar$. 

Physically, 
each resonance is associated with a {\em metastable state}: a
(not square-integrable) solution of the Schr\"odinger equation at the energy $z_j$,
which decays like $\E^{-t\gamma_j/\hbar}$ when $t\to+\infty$. 
In spectroscopy experiments, one measures the energy dependence of some scattering cross-section $\sigma(E)$.
Each resonance $z_j$ imposes a Lorentzian component 
$\frac{\gamma_j}{(E-E_j)^2+\gamma_j^2}$ on $\sigma(E)$; a resonance $z_j$
will be detectable on the signal $\sigma(E)$ only if its Lorentzian is 
well-separated from
the ones associated with nearby resonances of comparable widths, therefore iff
$|E_j'-E_j|\gg \gamma_j$. This condition of ``well-separability'' is NOT the one
we will be interested in here. We will rather consider the order of magnitude of each
resonance lifetime $\hbar/\gamma_j$, independently of the nearby ones, in the semiclassical r\'egime:
a resonant state will be ``visible'', or ``long-living'', if $\gamma_j=\Oo(\hbar)$. 
Our objective will be to {\rm count} 
the number of resonances $z_j$ in boxes of the type
$\{|E_j-E|\leq C\hbar\,,\,\gamma_j\leq C\hbar\}$, or equivalently $\{|z_j-E|\leq C\hbar\}$. 


\subsection{Trapped sets}
Since resonant states are ``invariant up to rescaling'', it
is natural to relate them, in the semiclassical spirit, to invariant structures of the
classical dynamics. For a scattering system, the set of points (of energy $E$)
which don't escape to infinity (either in past future) is 
called the \emph{trapped set} at energy $E$, and denoted by $K(E)$. The textbook
example of a radially-symmetric potential shows that this set may be empty (if $V(r)$ 
decreases monotonically from $r=0$ to $r\to\infty$), or have the same dimension as
$\Sigma_E$ (if $V(r)$ has a maximum $V(r_0)>0$ before decreasing as $r\to\infty$).

For $n=2$ degrees of freedom, the geometry of the trapped set can be more complex. 
Let us consider the 
well-known example of $2$-dimensional scattering by a set of non-overlapping disks 
\cite{GasRic,Cv-E} (a similar model was studied in \cite{Troll,BluSmi}). 

When the scatterer is a single disk, the trapped set is obviously empty. 

The scattering by two disks admits a single trapped periodic
orbit, bouncing back and forth between the disks.  Since the evolution between two bounces is ``trivial'',
it is convenient to represent the scattering system
through the \emph{bounce map} on the reduced
phase space (position along the boundaries $\times$ velocity angle). This map 
is actually defined only on
a fraction of this phase space, namely on those points which will bounce again at least once.
For the 2-disk system, this map
has a unique periodic point (of period $2$), which is of hyperbolic
nature due to the curvature of the disks. The trapped set $K$ of the map 
(``reduced'' trapped set)
reduces to this pair of
points; it lies at the intersection of the forward trapped set $\Gamma_-$ 
(points trapped as $t\to +\infty$) and the
backward trapped set $\Gamma_+$ (points trapped as $t\to -\infty$).

The addition of a third disk generates a 
complex bouncing dynamics, for which the trapped set is a 
\emph{fractal repeller} \cite{GasRic}. We will explain in the next section 
how such a structure arises
in the case of the open baker's map.
As in the $2$-disk case, the bounce map
is uniformly hyperbolic; each forward trapped point $x\in \Gamma_-$ admits a
stable manifold $W_-(x)$ (and vice-versa for $x\in\Gamma_+$). 
One can show that $\Gamma_-$ is fractal along the unstable direction $W_+$:
$\Gamma_-\cap W_+$ has a Hausdorff dimension $0<d<1$ which depends on the positions and sizes of
the disks. Due to time-reversal symmetry,
$\Gamma_+\cap W_-$ has
the same Hausdorff dimension. Finally, the reduced trapped set $K=\Gamma_+\cap\Gamma_-$
is a fractal of dimension $2d$, which contains infinitely many periodic orbits. 
The unreduced trapped set $K(E)\subset\Sigma_E$ has one more dimension corresponding 
to the direction of the flow,
so it is of dimension $D=2d+1$.

\subsection{Fractal Weyl law}
We now relate the geometry of the trapped set $K(E)$, to the density of resonances
of the quantized Hamiltonian $H_h$ in boxes $\set{|z-E|\leq C\hbar}$.
The following conjecture (which dates back at least to the work of Sj\"ostrand \cite{SjDuke})
relates this density with the ``thickness'' of the trapped set. 
\begin{conjecture}\label{conj:fWl}
Assume that the trapped set $K(E)$ at energy $E$ has dimension
$2d_E+1$. Then, the density of resonances near $E$
grows as follows in the semiclassical limit:
\begin{equation}
\label{eq:fweylh}
\forall r>0,\quad
\frac{\#\big\{ \mathrm{Res}(H_\hbar) \cap \set{ z \; : \; \ |z - E | < r\,\hbar  } \big\}}{\hbar^{-d_E}}\hto0 
c_E(r)\,,
\end{equation}
for a certain ``shape function'' $0\leq c_E(r)<\infty$. 
\end{conjecture}
We were voluntarily rather vague on the concept of ``dimension'' (a fractal set
can be characterized by many different dimensions).
In the case of a closed system, $K(E)$ has dimension
$2n-1$, so we recover the Weyl law (\ref{eq:weyl-closed}). 
If $K(E)$ consists in one
unstable periodic orbit, the resonances form a (slightly deformed)
rectangular lattice of sides $\propto \hbar$, so each $\hbar$-box 
contains at most finitely many resonances \cite{Sj2}.

For intermediate situations ($0<d_E<n-1$), one has only been able to prove
one half of the above estimate, namely the \emph{upper bound} for this resonance counting
\cite{SjDuke,ZwIn,GLZ,SjZw04}. The dimension appearing in these upper bounds is
the \emph{Minkowski dimension} defined by measuring $\ep$-neighborhoods of $K(E)$.
In the case we will study, this dimension is equal to the Hausdorff one.
Some lower bounds for the resonance density have been obtained as well \cite{SjZw-lower}, but
are far below the conjectured estimate.

Several numerical studies have
attempted to confirm the above estimate for a variety of scattering 
Hamiltonians \cite{GLZ,L,LZ,LSZ}, but with rather inconclusive results.
Indeed, it is numerically demanding to compute resonances. One method is to
``complex rotate'' the original Hamiltonian into a non-Hermitian operator, the eigenvalues
of which are the resonances. 
Another method uses the (approximate) 
relationship between, on one side, the resonance spectrum
of $H_\hbar$, one the other side, the set of zeros of some
semiclassical zeta function, which is computed from the knowledge of classical 
periodic orbits \cite{Cv-E,LSZ}. 
In the case of the geodesic flow on a convex
co-compact quotient of the Poincar\'e disk (which has a fractal trapped set), 
the resonances of the Laplace operator
are \emph{exactly} given by the zeros of Selberg's zeta function. Even in that case,
it has been difficult to check the asymptotic Weyl law (\ref{eq:fweylh}), due to the necessity
to reach sufficiently high values of the energy \cite{GLZ}.

\subsection{Open maps}
Confronted with these difficulties to deal with open Hamiltonian systems, we decided
to study semiclassical resonance distributions for
toy models which have already proven efficient to modelize closed systems. 
In the above example of obstacle scattering, the {\em bounce
map} emerged as a way to simplify the description of the classical dynamics. 
It acts on a reduced phase space,
and gets rid of the ``trivial'' evolution between bounces. The exact
quantum problem also reduces to analyzing an operator acting on wavefunctions
on the disk boundaries, but this operator is infinite-dimensional,
and extracting its resonances is not a simple task \cite{GasRic, BluSmi}.

Canonical maps on the $2$-torus were often used to mimic 
closed Hamiltonian systems; they can be quantized into unitary matrices,  
the eigenphases of which are to be compared with the eigenvalues $\E^{-\I E_j/\hbar}$ of the
propagator $\E^{-\I H_\hbar/\hbar}$ (see e.g. \cite{DEGra} and references therein for
a mathematical introduction on quantum maps). 

We therefore decided to construct a ``toy bounce map'' on $\t2$, 
with dynamics similar to the 
original bounce map, and which can be easily quantized
into an $N\times N$ {\em subunitary matrix} (where $N=(2\pi\hbar)^{-1}$). This matrix
is then easily diagonalized, and its subunitary eigenvalues
$\{\lambda_j\}$
should be compared with the set $\{\E^{-\I z_j/\hbar}\}$, where the $z_j$ are
the resonances of $H_\hbar$ near some positive energy $E$. 
We cannot prove any direct correspondence between, on one side the eigenvalues of
our quantized map, on the other side resonances of a {\em bona fide} scattering
Hamiltonian. However, we expect a semiclassical property like the fractal Weyl law
to be {\em robust}, in the sense that it should be shared by all types of 
``quantum models''. To support this claim, we notice that the usual, 
``closed'' Weyl law is already (trivially) satisfied by
quantized maps: the number of eigenphases $\theta_j$ on the unit circle (corresponding to
an energy range $\Delta E=2\pi\hbar$) is exactly $N=(2\pi\hbar)^{-1}$, which agrees with
the Weyl law (\ref{eq:weyl-closed}) for $n=2$ degrees of freedom.
Testing Conjecture \ref{conj:fWl} in the framework of quantum maps should therefore
give a reliable hint on its validity for more realistic Hamiltonian
systems. 

Schomerus and Tworzyd{\l}o recently studied the quantum spectrum of an open chaotic map
on the torus, namely the open kicked rotator \cite{schomerus}; they obtain a good
agreement with a fractal Weyl law for the resonances (despite the fact that the geometry
of the trapped set is not completely understood for that map).
The authors also provide a heuristic argument to explain this Weyl law.
We believe that this argument, upon some technical improvement, could yield a rigorous proof
of the upper bound for the fractal Weyl law in case of maps.

We preferred to investigate that problem using one of the best understood
chaotic maps on $\t2$, namely the baker's map.


\section{The open baker's map and its quantization}

\subsection{Classical closed baker}
The (closed) baker's map is one of the simplest examples of uniformly hyperbolic, strongly 
chaotic systems (it is a perfect model of Smale's horseshoe).
The ``3-baker's map'' $B$ on $\t2\equiv [0,1)\times [0,1)$ is defined as follows:
\bequ\label{e:closedB}
\t2\ni(q,p)\mapsto B(q,p)=\left\{\begin{array}{ll}(3q,\frac{p}{3})&{\rm if}\ 0\leq q<1/3,\\
(3q-1,\frac{p+1}{3})&{\rm if}\ 1/3\leq q<2/3,\\
(3q-2,\frac{p+2}{3})&{\rm if}\ 2/3\leq q<1.\end{array}\right.
\end{equation}
This map preserves the symplectic form $dq\wedge dp$ on $\t2$, and is invertible.
Compared with a generic Anosov map, it has the particularity 
to be linear by parts, and its linearized dynamics (well-defined away from its
lines of discontinuities) is independent of the point $x\in\t2$.  
As a consequence, the stretching exponent is constant on $\t2$, as well
as the unstable/stable directions (horizontal/vertical).

This map admits a very simple Markov partition, made of the three vertical rectangles
$R_j=\{q\in [j/3,(j+1)/3),\,p\in[0,1)\}$, $j=0,1,2$ (see Fig.~\ref{f:C}). 
Any bi-infinite sequence of symbols
$\ldots\ep_{-2}\ep_{-1}\cdot \ep_0\ep_1\ep_2\ldots$ (where each $\ep_i\in\{0,1,2\}$) will
be associated with the \emph{unique} point $x$ s.t. $B^t(x)\in R_{\ep_t}$ for all $t\in\ZZ$. This 
is the point of coordinates
$(q,p)$, where $q$ and $p$ admit the ternary decompositions 
$$
q=0\cdot\ep_0\ep_1\ldots\defeq\sum_{i\geq 1} \frac{\ep_{i-1}}{3^{i}}\,,\qquad p=0\cdot\ep_{-1}\ep_{-2}\ldots\,.
$$
The baker's map $B$ simply acts
as a {\em shift} on this symbolic sequence:
\bequ\label{e:shift}
B(x=\ldots\ep_{-2}\ep_{-1}\cdot \ep_0\ep_1\ep_2\ldots)=\ldots \ep_{-2}\ep_{-1}\ep_0\cdot\ep_1\ep_2\ldots\,.
\end{equation}

\subsection{Opening the classical map\label{s:open}}
We explained above that the bounce maps associated with the $2$- or $3$-disk systems
were defined only on parts of the reduced phase space,
namely on those points which bounce at least one more time. The remaining points,
which escape to infinity right after the bounce, have no image through the map.

\begin{figure}[ht]
\begin{center}
\includegraphics[height=4.5cm]{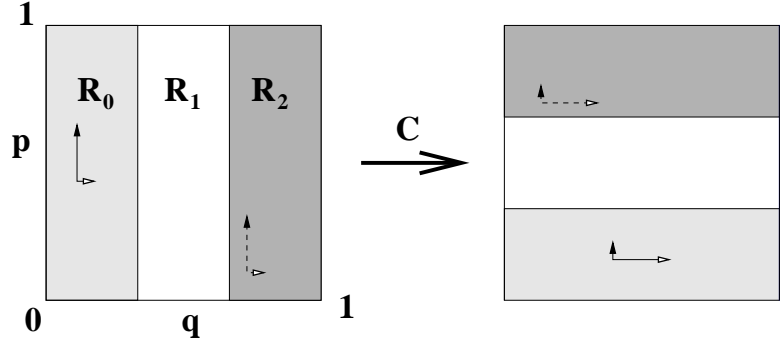}
\end{center}
\caption{\label{f:C}Open baker's map $C$. The points in the middle rectangle are sent to infinity.}
\end{figure}

\begin{figure}[htbp]
\begin{center}
\includegraphics[height=5cm]{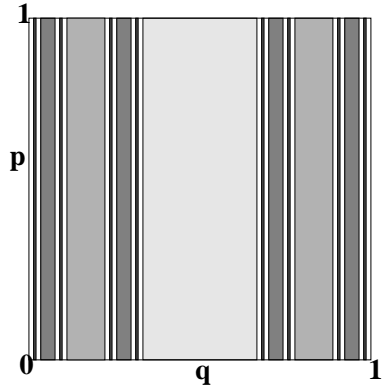}
\end{center}
\caption{\label{f:Gamma-}Iterative construction of the forward trapped set $\Gamma_-$ 
for the open baker's map $C$: we
remove from $\t2$ the points leaving to infinity at times $t=1,\,2,\,3,\,4$ 
(from light grey to dark grey) etc. At the end, there remains the fractal set 
$\Gamma_-=Can\times[0,1)$.}
\end{figure}

Hence, to open our baker's map $B$, we just decide to restrict it
on a subset $S\subset \t2$, or equivalently we send points in $\t2\setminus S$
to infinity.
We obtain an Anosov map ``with a hole'', a class of dynamical systems 
recently studied in the literature \cite{Cher2}. The study is simpler when the hole
corresponds to a Markov rectangle \cite{Cher1}, so this is 
the choice we will make (we expect the fractal Weyl law to hold
for an arbitrary hole as well). 
Let us choose for the hole the second Markov rectangle $R_1$, so that $S=R_0\cup R_2$.
Our open map $C=B_{\restriction S}$ reads (see Fig.~\ref{f:C}):
\bequ\label{eq:C}
C(q,p)=\left\{\begin{array}{ll}(3q,\frac{p}{3})&{\rm if}\ q\in R_0,\\
(3q-2,\frac{p+2}{3})&{\rm if}\ q\in R_2.\end{array}\right.
\end{equation}
This map is canonical on $S$, and its inverse $C^{-1}$ 
is defined on the set $C(S)$. 
Our choice for $S$ coincides with the points $x=(q,p)$
satisfying $\ep_0(x)\in\{0,2\}$ (equivalently, points
s.t. $\ep_0(x)=1$ are sent to infinity through $C$). 
This allows us to characterize the trapped sets very easily:
\begin{itemize}
\item the forward trapped set $\Gamma_-$ (see fig.~\ref{f:Gamma-}) is made of the
points $x$ which will never fall
in the strip $R_1$ for times $t\geq 0$: these are the points s.t. 
$\ep_i\in\{0,2\}$ for all $i\geq 0$, with no constraint on the $\ep_i$ for $i<0$.
This set is of the form $\Gamma_-=Can\times [0,1)$, where $Can$ is the
standard $1/3$-Cantor set on the unit interval. As a result, the intersection $\Gamma_-\cap W_+\equiv Can$
has the Hausdorff (or Minkowski) dimension 
$d=\frac{\log 2}{\log 3}$.
\item the backward trapped set $\Gamma_+$ is made of the points satisfying 
$\ep_i\in\{0,2\}$ for all $i<0$, and is given by $[0,1)\times Can$.
\item the full trapped set $K=Can\times Can$.
\end{itemize}

\subsection{Quantum baker's map}
We now describe in some detail the quantization of the above maps.
We recall \cite{DEGra,DBgiens} that a nontrivial quantum Hilbert space
can be associated with the phase space $\t2$ only for discrete values of 
Planck's constant, namely $\hbar=(2\pi N)^{-1}$, $N\in\NN_0$. In that case
(the only one we will consider),
this space $\hn$ is of dimension $N$. It admits the ``position'' basis 
$\{Q_j,\ j=0,\ldots,N-1\}$ made of the ``Dirac combs'' 
$$
Q_j(q)=\frac{1}{\sqrt{N}}\sum_{\nu\in\ZZ} \delta(q-\frac{j}{N}-\nu)\,.
$$
This basis is connected to the ``momentum'' basis $\{P_k,\,k=0,\ldots,N-1\}$ through
the discrete Fourier transform:
\bequ\label{e:FT}
\langle P_k|Q_j\ra=(\F_N)_{kj}=\frac{\E^{-2\I\pi N kj}}{\sqrt{N}}\,,\quad j,k\in\{0,\ldots,N-1\}\,,
\end{equation}
where the Fourier matrix $F_N$ is unitary. Balazs and Voros \cite{BaVo} proposed to
quantize the closed baker's map $B$ as follows, when $N$ 
is a multiple of $3$ (a condition we will always assume):
in the position basis, it takes the block form
\bequ
B_N=\F_N^{-1}\left(\begin{array}{ccc}\F_{N/3}&&\\&\F_{N/3}&\\&&\F_{N/3}\end{array}\right)\,.
\end{equation}
This matrix is obviously unitary, and exactly satisfies the Van Vleck formula
(the semiclassical expression for a quantum propagator, in terms of the 
classical generating function). In the semiclassical limit $N\to\infty$,
it was shown \cite{DENW} that these matrices classically propagate Gaussian coherent states
supported far enough from the lines of discontinuities. As usual,
discontinuities of the classical dynamics induce diffraction effects at the quantum level,
which have been partially analyzed for the baker's map \cite{ToVaSa} (in particular,
diffractive orbits have to be taken into account in the Gutzwiller formula for 
$\mathrm{tr}(B_N^t)$). We believe that these diffractive effects should only induce lower-order
corrections to the Weyl law (\ref{e:fWl2}).

\medskip

\begin{figure}[htbp]
\begin{center}
\rotatebox{-90}{\includegraphics[height=11cm]{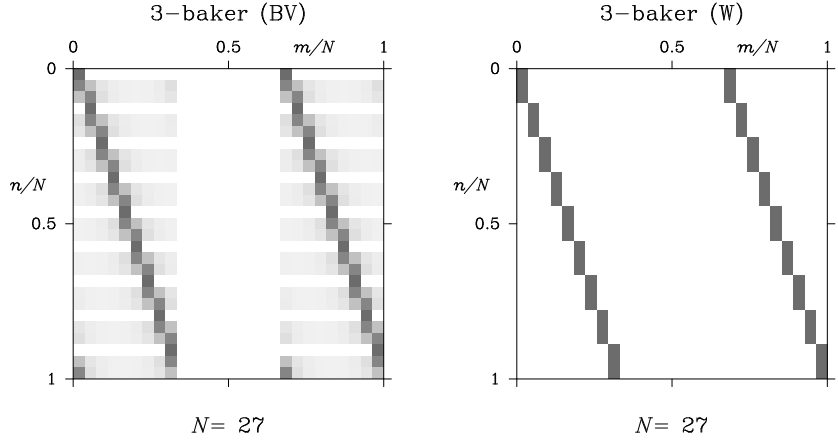}}
\end{center}
\caption{\label{f:matrices}Graphical representation of the matrices $C_N$ 
(\ref{e:C_N}) and $\widetilde C_N$ (\ref{e:tC9}). 
Each grey square represents the modulus of a matrix
element (white=$0$, black=$1/\sqrt{3}$)}
\end{figure}

We are now ready to quantize our open baker's map $C$ of (\ref{eq:C}): since the classical map
sends points in $R_1$ to infinity and acts through $B$ on $S=R_0\cup R_2$, the quantum 
propagator should kill states microsupported on $R_1$, and act as $B_N$ on states
microsupported on $S$. Therefore, in the position basis we get the subunitary matrix
\bequ\label{e:C_N}
C_N=\F_N^{-1}\left( \begin{array}{ccc}\F_{N/3}&&\\&0&\\&&\F_{N/3}\end{array}\right)\,.
\end{equation}
A very similar open quantum baker was constructed in \cite{SaVa}, as a quantization of Smale's horseshoe.
In Figure~\ref{f:matrices} (left) we represent the moduli of the matrix elements $(C_N)_{nm}$. The largest
elements are situated along the ``tilted diagonals'' $n=3m$, $n=3(m-2N/3)$, which correspond to the projection on
the $q$-axis of the graph of $C$. Away from these
``diagonals'', the amplitudes of the elements decrease relatively slowly (namely, like $1/|n-3m|$). This slow
decrease is due to the diffraction effects associated with the discontinuities of the map.
\subsection{Resonances of the open baker's map}
\begin{figure}[htbp]
\begin{center}
\includegraphics[width=11cm]{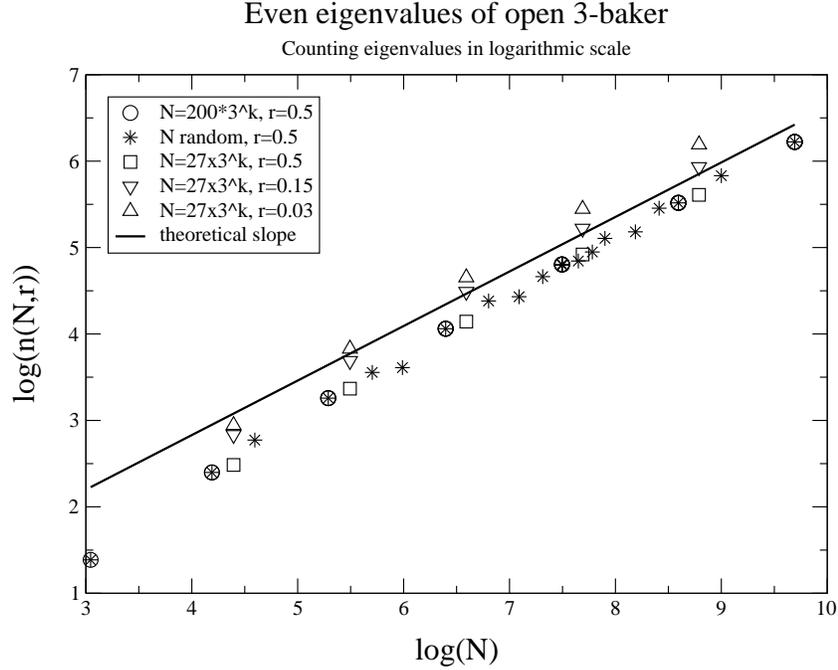}
\caption{\label{f:table3e} Checking the $N$-dependence of $n(N,r)$ for various
values of $r$, along geometric and arbitrary sequences for $N$. The thick curve has
the slope $\log2/\log 3$}
\end{center}
\end{figure}

We numerically diagonalized the matrices $C_N$, for larger and larger Planck's constants $N$.
First of all, we notice that the subspace $\mathrm{Span}\{Q_j\,,\ j=N/3,\ldots, 2N/3-1\}$,
made of position states in the ``hole'', is in the kernel of $C_N$. Therefore, it is sufficient to diagonalize the 
matrix obtained by removing the corresponding lines and columns. 
Upon a slight modification of the quantization 
procedure \cite{Sa},
one obtains for $C_N$ a matrix covariant w.r.to parity, allowing for a separation of the even and odd
eigenstates, and therefore reducing the dimension of each part by $2$. This is the quantization
we used for our numerics: we only plot the even-parity resonances
(the distribution of the odd-dimensional ones is very similar). 
\begin{figure}[htbp]
\begin{center}
\rotatebox{-90}{\includegraphics[width=6cm]{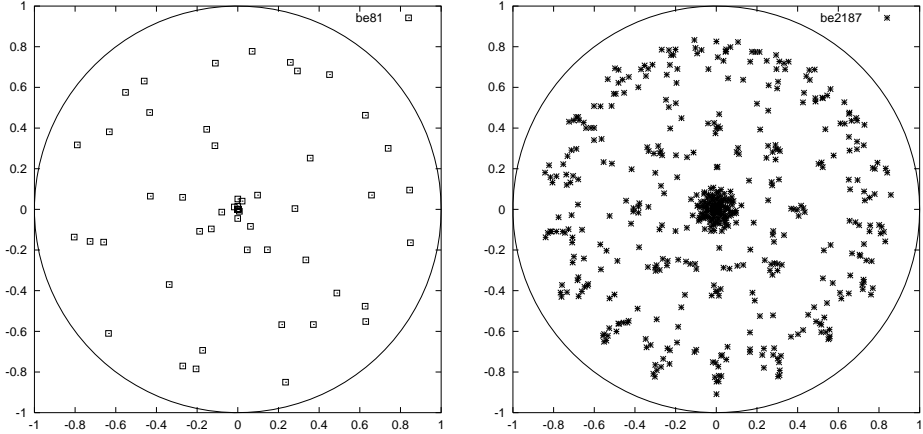}}
\end{center}
\caption{\label{f:spect3be} Even-parity spectrum of the matrices $C_N$
for $N/3=81,\,2187$}
\end{figure}
In figure~\ref{f:spect3be} we show the even-parity
spectra of the matrix $C_N$ for $N=3^5$ and $N=3^8$. 
Although we could 
not detect exact null states for the reduced matrix, many among the $N/3$ eigenvalues had very small
moduli: for large values of $N$, the spectrum of $C_N$ accumulates near the origin.
This accumulation is an obvious consequence of the fractal Weyl law we want to test:
\begin{conjecture}
For any radius $1>r>0$ and $N\in\NN_0$, $3|N$, let us denote 
$$
n(N,r)\defeq\#\big\{ \lambda\in\mathrm{Spec}(C_N) \cap \set{ |\lambda|\geq r  } \big\}\,.
$$
In the semiclassical limit, this counting function behaves as
\bequ\label{e:fWl2}
\frac{n(N,r)}{N^{\frac{\log 2}{\log 3}}}\Nto8 c(r)\,,
\end{equation}
with a ``shape function'' $0\leq c(r)<\infty$.
\end{conjecture}
To test this conjecture, we proceed in two ways:

$\bullet$ In a first step, we select some discrete values for $r$, and plot $n(N,r)$ for
an arbitrary sequence of $N$, in a log-log plot (see Fig.~\ref{f:table3e}). 
We observe
that the slope of the data nicely converges towards the theoretical one $\frac{\log 2}{\log 3}$ (thick line), 
all the more so along geometric subsequences $N=3^k\,N_o$,
and for relatively large values of the radius ($r=0.5$). For the smaller value
$r=0.03$, the annulus $\{|z|\geq r\}$ still contains ``too many resonances'' and the asymptotic r\'egime
is not yet reached.

\begin{figure}[tbp]
\includegraphics[width=12cm]{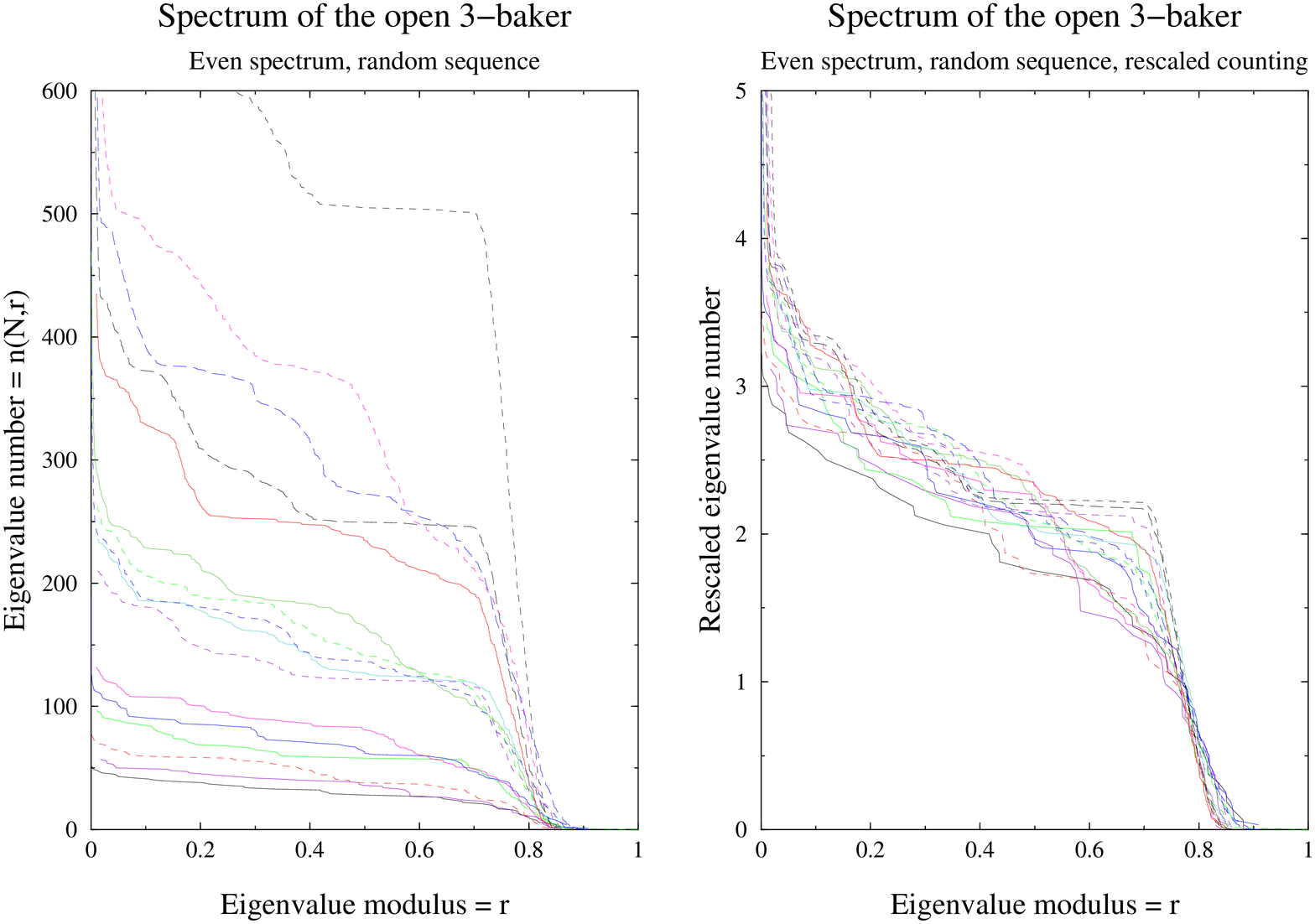}
    \caption{\label{f:spect-histo3be} On the left, we plot the number $n(N,r)$ of 
      even eigenvalues of $C_N$ of modulus $\geq r$. On  the
      right plot, we rescale those functions by the factors $N^{-\frac{\log 2}{\log 3}}$}
\end{figure}

$\bullet$ In a second step, confident that $n(N,r)$ scales like $N^{\frac{\log 2}{\log 3}}$, 
we try to extract the shape function $c(r)$.
For an arbitrary sequence of values of $N$, we plot the function $n(N,r)$ 
(Fig.~\ref{f:spect-histo3be}, left), and then rescale the
vertical coordinate by a factor $N^{-\frac{\log 2}{\log 3}}$ (right).
The rescaled curves do roughly superpose on one another, supporting
the conjecture. However, there remains relatively large fluctuations, even for large values of $N$.
The curves corresponding to a {\em geometric sequence} $N=3^k\,N_o$, $k=0,1,\ldots$ tend to be nicely
superposed to one another, but slightly differ from one sequence to another. 
Similar plots were given in \cite{schomerus} in the case of the kicked rotator; the shape function
$c(r)$ is conjectured there to correspond to some ensemble of random subunitary matrices. Our data are
too unprecise to perform such a check.

The fact that the spectra of the matrices $C_N$ ``behave nicely'' along geometric sequences, 
while they fluctuate
more strongly between successive values of $N$, is not totally unexpected 
(similar phenomena had been noticed for the
quantizations $B_N$ of the closed baker \cite{BaVo}). 
In view of Fig.~\ref{f:spect-histo3be}, our conjecture (\ref{e:fWl2}) may be too strict if
we apply it to a general sequence of $N$. At least, it seems to be satisfied along 
geometric sequences $\set{3^k\,N_o,\,k\in\NN}$, 
with shape functions $c_{N_o}(r)$ slightly depending on the sequence.

\section{A solvable toy model for the quantum baker}

\subsection{Description of the toy model}
In an attempt to get some analytical grip on the resonances, we tried to simplify the quantum matrix
$C_N$, keeping only its ``backbone'' along the tilted diagonals and removing the off-diagonal components. 
We obtained the ``toy-of-the-toy model'' given by the following matrices (the moduli of the components
are shown on right plot of Fig.~\ref{f:matrices}):
\begin{equation}
    \label{e:tC9}
    \tC_{N=9} = \frac{1}{\sqrt 3} \left( \begin{array}{lllllllll}
      1 & 0 & 0 & 0 & 0 & 0 & 1 & 0 & 0 \\
      1 & 0 & 0 & 0 & 0 & 0 & \omega^2 & 0 & 0 \\
      1 & 0 & 0 & 0 & 0 & 0 & \omega & 0 & 0 \\
      0 & 1 & 0 & 0 & 0 & 0 & 0 & 1 & 0 \\
      0 & 1 & 0 & 0 & 0 & 0 & 0 & \omega^2 & 0 \\
      0 & 1 & 0 & 0 & 0 & 0 & 0 & \omega & 0 \\
      0 & 0 & 1 & 0 & 0 & 0 & 0 & 0 & 1 \\
      0 & 0 & 1 & 0 & 0 & 0 & 0 & 0 & \omega^2 \\
      0 & 0 & 1 & 0 & 0 & 0 & 0 & 0 & \omega 
    \end{array} \right), \quad \omega = \E^{ 2 \pi \I / 3 }\,.
\end{equation}
From this example, it is pretty clear how one constructs $\widetilde C_N$ for $N$ an 
arbitrary multiple of $3$. A similar quantization
of the closed $2$-baker was introduced in \cite{schack}.

Before describing the spectra of these matrices, we describe their propagation properties.
Removing the ``off-diagonal'' elements, we have eliminated the effects of diffraction due to the
discontinuities of $C$. However, this elimination is so abrupt that it modifies
the semiclassical transport. Indeed, a coherent state situated at a
point $x$ away from the discontinuities will not be transformed by $\tC_N$ into a single coherent state 
(as does $C_N$),
but rather into a {\em linear combination} of $3$ coherent states, shifted vertically by $1/3$ from one another. 
Therefore, the matrices $\tC_N$ do not quantize the open baker $C$ of (\ref{eq:C}),
but rather the following {\em multivalued} (``ray-splitting'') map:
\bequ
\tC(q,p)=\left\{\begin{array}{ll}(3q,\frac{p}{3})\cup(3q,\frac{p+1}{3})\cup(3q,\frac{p+2}{3})&{\rm if}\ q\in R_0,\\
(3q-2,\frac{p}{3})\cup(3q-2,\frac{p+1}{3})\cup(3q-2,\frac{p+2}{3})&{\rm if}\ q\in R_2.\end{array}\right.
\end{equation}
This modification of the classical dynamics is rather annoying. Still, the dynamics $\tC$ 
shares some common features with that of $C$: the forward trapped set for $\tC$ is the same
as for $C$, that is the set $\Gamma_-$ described in Fig.~\ref{f:Gamma-}.
On the other hand, the backward trapped set is now the full torus $\t2$. 

\subsection{Interpretation of $\tC_N$ as a Walsh-quantized baker}
A possible way to avoid this modified classical dynamics is to interpret
$\tC_N$ as a ``Walsh-quantized map'' (this interpretation makes sense when $N=3^k$, $k\in\NN$). 
To introduce this Walsh
formalism, let us first write the Hilbert space as a tensor product $\Hh_N=(\CC^3)^{\otimes k}$, 
where we take the ternary decomposition of discrete positions
$\frac{j}{N}=0\cdot \ep_0\ep_1\cdots\ep_{k-1}$ into account. 
If we call $\{e_0,\ e_1,\ e_2\}$ the canonical basis
of $\CC^3$, each position state $Q_j\in\Hh_N$ can be represented as the tensor product state
$$
Q_j=e_{\ep_0}\otimes e_{\ep_1}\otimes\cdots\otimes e_{\ep_{k-1}}\,.
$$
In the language of quantum computing, each tensor factor $\CC^3$ is the Hilbert space of 
a ``qutrit'' associated with a certain scale \cite{schack}.

The Walsh Fourier transform is a modification of the discrete Fourier transform (\ref{e:FT}), 
which first appeared in
signal theory, and has been recently used as a toy model for harmonic analysis \cite{muscalu}. Its major
advantage is the possibility to construct states compactly supported in both position and ``Walsh momentum''.
In our finite-dimensional framework, this Walsh transform is the matrix
$$
(W_N)_{j j'}=3^{-k/2}\,\exp\Big(-\frac{2i\pi}{3} \sum_{\ell+\ell'=k-1} \ep_\ell(Q_j)\,\ep_{\ell'}(Q_j')\Big)\,,
\qquad j,j'=0,\ldots, N-1\,,
$$
and acts as follows on tensor product states: 
$$
W_N\,(v_0\otimes v_1\otimes\cdots v_{k-1})=\F_3 v_{k-1}\otimes \cdots \F_3 v_{1}\otimes\F_3 v_{0}\,,
\quad v_{\ell}\in\CC^3,\ \ell=0,\ldots,k-1\,.
$$
Now, in the case $N=3^k$, our toy model $\tC_N$ can be expressed as
$$
\tC_N=W_N^{-1}\left(\begin{array}{ccc}W_{N/3}&&\\&0&\\&&W_{N/3}\end{array}\right)\,.
$$
One can show that ``Walsh coherent states'' are propagated through $\tC_N$ according
to the map $C$. Hence, 
as opposed to what happens in ``standard'' quantum mechanics, 
$\tC_N$ Walsh-quantizes the open baker $C$.

\subsection{Resonances of $\tC_{N=3^k}$}
We now use the very peculiar properties of the matrices $\tC_{3^k}$
to analytically compute their spectra.
\begin{figure}[htbp]
\begin{center}
\rotatebox{-90}{\includegraphics[width=8cm]{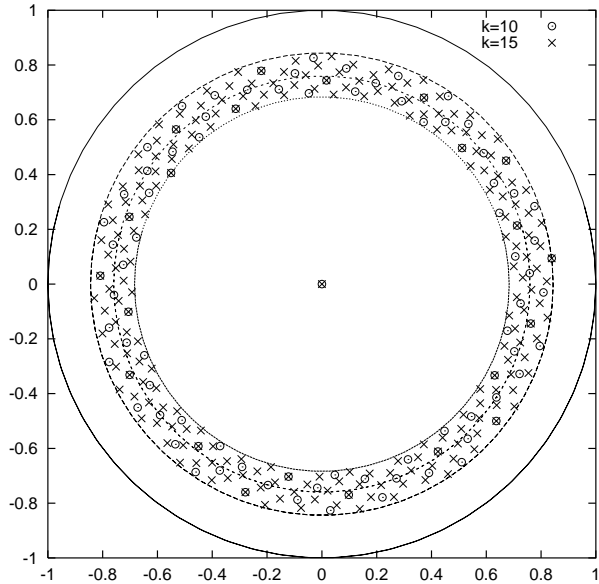}}
\end{center}
\caption{\label{f:toy} Spectra of the matrices $\tC_N$
for $N=3^{10}$ (circles), $N=3^{15}$ (crosses). The large circles have radii $|\lambda|=1,\,|\lambda_+|,\,
| \lambda_+ \lambda_-|^{\frac12},\, |\lambda_-| $}
\end{figure}
From the expressions in last section, one can see that the toy model $\tC_N$ acts 
very simply on tensor product states:
\begin{equation}
\tC_N\,v_0\otimes v_1\otimes\cdots v_{k-1}=v_1\otimes\cdots v_{k-1}\otimes \F^{-1}_3\pi_{02} v_0\,,
\end{equation}
where $\pi_{02}$ projects $\CC^3$ orthogonally onto $\mathrm{Span}\set{e_0,\,e_2}$.
Like its classical counterpart,
$\tC_N$ realizes a symbolic shift between the different scales. 
It also sends the first symbol $\ep_0$ to the ``end of the queue'', after a projection
and a Fourier transform. The projection $\pi_{02}$ kills the states $Q_j$ localized
in the rectangle $R_1$. The vector $\F^{-1}_3 e_{\ep_0}$ in the last qutrit
induces a localization in the momentum
direction, near the momentum $p=0\cdot\ep_0$.

By iterating this expression $k$ times, we see that the operator $(\tC_N)^k$ acts independently on
each tensor factor $\CC^3$, through the matrix $\F^{-1}_3\pi_{02}$. The latter has three
eigenvalues: 
\begin{itemize}
\item it kills the state $e_{1}$, implying that $(\tC_N)^k$ kills any state $Q_j$ for which at 
least one of the symbols
$\ep_\ell(Q_j)$ is equal to $1$. These $3^k-2^k$ position states are localized
``outside'' of the trapped set $\Gamma_-$, which explains why they are killed by the dynamics.
\item its two remaining eigenvalues $\lambda_\pm$ have moduli 
$|\lambda_+|\approx 0.8443$, $|\lambda_-|\approx 0.6838$. They build up the ($2^k$-dimensional)
nontrivial spectrum of
$\tC_N$, which has the form of a ``lattice'' (see Fig.~\ref{f:toy}):
\end{itemize}
\begin{proposition}
For $N=3^k$, the nonzero spectrum of $\tC_N$ is the set
$$
\set{\lambda_+}\cup\set{\lambda_-}\bigcup \set{ \E^{2\I\pi\frac{j}{k}}\, \lambda_+^{1 - p/k} \lambda_-^{p/k}
\; : \; 1 \leq p \leq k-1\,,\ 0\leq j\leq k-1 } \,.
$$
Most of these eigenvalues are highly degenerate (they span a subspace of dimension $2^k$).
When $k\to\infty$, the highest degeneracies occur when $p/k\approx 2$,
which results in the following asymptotic distribution:
$$
\forall f\in C(\RR^2),\quad
\lim_{k\to\infty}\ \frac{1}{2^k}
\sum_{\lambda\in\mathrm{Spec}(\tC_{3^k})\setminus 0}\mathrm{mult}(\lambda)\,f(\lambda)=
\int_0^{2\pi} f(|\lambda_-\lambda_+|^{1/2},\theta)\,\frac{\D\theta}{2\pi}\,.
$$
\end{proposition} 
The last formula shows that the spectrum of $\tC_N$ 
along the geometric sequence $\{N=3^k,\,k\in\NN\}$ satisfies the fractal Weyl law (\ref{e:fWl2}), 
with a shape function in form of an abrupt step: $c(r)= \Theta(|\lambda_+\lambda_-|^{1/2}-r)$.
Although the above spectrum seems very nongeneric (lattice structure, singular shape function), 
it is the first example (to our
knowledge) of a quantum open system proven to satisfy the fractal Weyl law.

\medskip

\noindent{\sc Acknowledgments.} We benefited from insightful discussions with
Marcos Saraceno, Andr\'e Voros, Uzy Smilansky, Christof Thiele and Terry Tao. 
Part of the work was done while I was visiting M.~Zworski in UC Berkeley, 
supported by the grant DMS-0200732 of the National Science Foundation.


\begin{thebibliography}{XX}

\bibitem{BaVo}N.L.~Balazs and A.~Voros  {\em The quantized baker's transformation},
Ann.\ Phys. {\bf 190} (1989), 1--31.

\bibitem{BluSmi}R.~Bl\"umel and U.~Smilansky, {\em A simple model for chaotic
scattering}, Physica {\bf D 36} (1989), 111--136.

\bibitem{Cher1}
N.~Chernov and R.~Markarian, 
{\it Ergodic properties of Anosov maps with rectangular holes},
Boletim Sociedade Brasileira Matematica {\bf 28} (1997), 271--314.

\bibitem{Cher2}N. Chernov, R. Markarian and S.~Troubetzkoy, 
{\it Conditionally invariant measures for Anosov maps with small holes}, 
Ergod. Th. Dyn. Sys. {\bf 18} (1998), 1049--1073.

\bibitem{Cv-E}
P.~Cvitanovi\'{c} and B.~Eckhardt, {\it Periodic-orbit quantization of chaotic systems},
Phys. Rev. Lett. {\bf 63} (1989) 823--826

\bibitem{DBgiens}S.~De~Bi\`evre, {\it Recent results on quantum map eigenstates},
these Proceedings.

\bibitem{DEGra}
M.~Degli~Esposti and S.~Graffi, editors
{\it The mathematical aspects of quantum maps}, volume 618 of
Lecture Notes in Physics, Springer, 2003.

\bibitem{DENW} M.~Degli~Esposti, S.~Nonnenmacher and B.~Winn, {\it Quantum variance and
ergodicity for the baker's map}, to be published in Commun. Math. Phys. (2005), arXiv:math-ph/0412058.

\bibitem{GasRic}P. Gaspard and S.A. Rice, {\it Scattering from a classically chaotic
repellor}, J. Chem. Phys. {\bf 90} (1989), 2225--2241; ibid, {\it Semiclassical
quantization of the scattering from a classically chaotic repellor},
J. Chem. Phys. {\bf 90} (1989), 2242--54; ibid, {\it Exact quantization of the scattering
from a classically chaotic repellor}, J. Chem. Phys. {\bf 90} (1989), 2255--2262;
Errata, J. Chem. Phys. {\bf 91} (1989), 3279--3280.

\bibitem{GLZ} L. Guillop\'e, K. Lin, and M. Zworski, 
{\em The Selberg zeta function for convex co-compact Schottky groups,}
Comm. Math. Phys, {\bf 245}(2004), 149--176.

\bibitem{Ivr} V. Ivrii, {\it Microlocal Analysis and Precise Spectral 
Asymptotics}, Springer Verlag, 1998.

\bibitem{L} K. Lin, {\em Numerical study of quantum resonances in          
chaotic scattering}, J. Comp. Phys. {\bf 176}(2002), 295--329.

\bibitem{LZ} K. Lin and M. Zworski, {\em Quantum resonances in          
chaotic scattering}, Chem. Phys. Lett. {\bf 355}(2002), 201--205.

\bibitem{LSZ} W. Lu, S. Sridhar, and M. Zworski,
{\em Fractal Weyl laws for chaotic open systems}, 
Phys. Rev. Lett. {\bf 91}(2003), 154101.

\bibitem{muscalu} C. Muscalu, C. Thiele, and T. Tao,
{\em A Carleson-type theorem for a Cantor group model of the  Scattering Transform,}
Nonlinearity {\bf 16} (2003), 219--246.

\bibitem{nz}S. Nonnenmacher and M. Zworski, {\em Distribution of resonances for open quantum maps},
preprint, 2005.

\bibitem{Sa} 
M.~Saraceno, {\it Classical structures in the quantized baker transformation},
Ann.\ Phys.\ (NY) {\bf 199} (1990), 37--60.

\bibitem{SaVa} M.~Saraceno and R.O.~Vallejos, {\it The quantized D-transformation}, 
Chaos {\bf 6} (1996), 193--199.

\bibitem{schack}
R.~Schack and C.M.~Caves, {\it Shifts on a finite qubit string: a class of quantum baker's maps},
Appl. Algebra Engrg. Comm. Comput. {\bf 10} (2000) 305--310

\bibitem{schomerus}
H.~Schomerus and J.~Tworzyd{\l}o, {\it Quantum-to-classical crossover of
quasi-bound states in open quantum systems}, Phys. Rev. Lett. {\bf 93} (2004),
154102.

\bibitem{SjDuke} J. Sj\"ostrand,  {\em
Geometric bounds on the density of resonances for semiclassical problems},
{ Duke Math. J.}, {\bf 60} (1990), 1--57.

\bibitem{Sj2} J. Sj\"ostrand,  {\em Semiclassical resonances generated by a non-degenerate
critical point}, Lecture Notes in Math. {\bf 1256}, 402--429, Springer (Berlin), 1987.


\bibitem{SjZw-lower}J. Sj\"ostrand and M.~Zworski, {\em Lower bounds on the number of
scattering poles}, Commun. PDE {\bf 18} (1993), 847--857.

\bibitem{SjZw04} J. Sj\"ostrand and M. Zworski,
{\em Geometric bounds on the density of semiclassical resonances in small domains,}
in preparation (2005).

\bibitem{ToVaSa} F.~Toscano, R.O.~Vallejos and M.~Saraceno,
{\em Boundary conditions to the semiclassical traces of the baker's map}, Nonlinearity {\bf 10}
(1997), 965--978.

\bibitem{Troll}G.~Troll and U.~Smilansky, {\em A simple model for chaotic scattering},
Physica {\bf D 35} (1989), 34--64.

\bibitem{ZwIn} M. Zworski, {\em Dimension of the limit set and the density of
resonances for convex co-compact Riemann surfaces}, Inv. Math. {\bf 136} (1999),
353--409.

\end{thebibliography}
\end{document}